# Analyzing the effects of surface distribution of pores in cell electroporation for a cell membrane containing cholesterol

**Pratip Shil[1,3], Salil Bidaye[2], Pandit B. Vidyasagar[1]**

[1]Biophysics Laboratory, Department of Physics, University of Pune, Pune 411007, India
[2] Institute of Bioinformatics and Biotechnology, University of Pune, Pune 411007, India
[3] Author to whom any correspondence should be addressed.
Email: pratip@physics.unipune.ernet.in

**Abstract**
This paper presents a model and numerical analysis (simulations) of transmembrane potential induced in biological cell membrane under the influence of externally applied electric field (i.e., electroporation). This model differs from the established models of electroporation in two distinct ways. Firstly, it incorporates the presence of cholesterol (~20% mole-fraction) in biological membrane. Secondly, it considers the distribution of pores as a function of the variation of $\phi_m$ from one region of the cell to another. Formulation is based on the role of membrane tension and electrical forces in the formation of pores in a cell membrane, which is considered as an infinitesimally thin insulator. The model has been used to explore the process of creation and evolution of pores and to determine the number and size of pores as a function of applied electric field (magnitude and duration). Results show that the presence of cholesterol enhances poration by changing the membrane tension. Analyses indicate that the number of pores and average pore radii differ significantly from one part of the cell to the other. While some regions of the cell membrane undergo rapid and dense poration, others remain unaffected. The method can be a useful tool for a more realistic prediction of pore formation in cells subjected to electroporation.

## 1. Introduction

Electroporation or electropermeabilization is a well known physical process in biological cells involving enhanced permeability of the cell membrane induced by an externally applied electric field [1-3]. The application of controlled high voltage pulses to the cell membrane leads to the formation of micro-pores in the cell membrane. This enables exogenous molecules viz drugs and DNA to move into the cells. Thus numerous applications in molecular biology, biotechnology and medicine have emerged [4-10].

Considerable efforts have also been made to understand this process and its optimization for various applications. It is established that the electropermeabilization of cells depends on the electric pulse amplitude, pulse duration, the number of pulses used and also on the experimental conditions such as buffer and temperature. With these parameters chosen, the process of permeabilization is reversible and the cell comes back to the normal physiological state. The process of electroporation is a two phase process: first phase is of permeabilization during the pulse, and second is a longer phase of resealing that begins after the end of the pulse [11]. A biological cell can be assumed to be a non-conducting sphere with a equipotential inner side (i.e, the cell membrane is an infinitesimally thin insulator). When a stepwise uniform electric field of magnitude $E_e$ is applied to a spherical cell of radius $R_{cell}$ in solution, the induced membrane potential is given by

$$\phi_m = 1.5 R_{cell} E_e \cos\theta \qquad (1)$$

where $\theta$ is the angle [1,2] as shown by Figure1(a). This equation states that at $\theta=0$ and at $\theta=\pi$ (referred to as 'poles'), the value of $\phi_m$ is extremum and the value is dependant on the cell size.

Also, the value of $\phi_m$ changes with $\theta$, meaning that different regions of the cell will experience different $\phi_m$.

Though reports of theoretical models and simulations of cell electroporation exist [12-14], major limitations remain as most of the models consider various approximations such as uniform poration of whole bilayer membrane. The effective models and numerical analysis (simulations) proposed by Neu and Krassowska and Smith *et al* provide considerable insight into the understanding of electroporation process [15-16].

An earlier report published by Smith *et al* [16] describes formation and evolution of pores and also proposes a model for DNA uptake mediated by electroporation which favoured experimental observations. However, the major limitations of this model involve: i) the assumption that the composition of the membrane is uniform in terms of membrane lipids (which does not take into account the presence of cholesterol) and ii) not considering the fact that the different regions of the cell experience different values of $\phi_m$ and hence has different pore densities (surface distribution of pores).

This work is aimed at overcoming these limitations and producing simulation for electroporation of mammalian cell membrane with a cholesterol content of 20% mole-fraction assumed to be distributed uniformly in the membrane.

**2. Calculation method**

According to the theory of electroporation, [1-3, 16] all the pores formed are initially hydrophobic and are formed at a rate determined by their energy. Most of them are quickly destroyed by lipid fluctuations, but if hydrophobic pores of radius $r_p > r^*$ ( where $r^*$ is critical radius) are created then they may be converted spontaneously into long-lived hydrophilic pores. Thus, most of the hydrophilic pores are initially created with radius slightly greater than $r^*$ and immediately expand to the minimum energy radius $r_m$. In the earlier model [16] for simulation studies, it has been assumed that the pores are created with initial radius $r_m$ ($r_p > r_m$). It has also been proposed that for any $r_p > r^*$, the pores will grow. The rate of creation of pores is given by,

$$\frac{dN}{dt} = \alpha e^{\left(\phi_m / \phi_{ep}\right)^2} \left(1 - \frac{N}{N_{eq}(\phi_m)}\right) \qquad (2)$$

where $N$ is the pore density, $\phi_m$ is the transmembrane potential, $\phi_{ep}$ is the characteristic voltage of electroporation, $\alpha = 1 \times 10^9$ m$^{-2}$s$^{-1}$ is the creation rate coefficient, $N_{eq}$ is the equilibrium pore density for a given voltage and is given by :

$$N_{eq}(\phi_m) = N_0 e^{q\left(\phi_m / \phi_{ep}\right)^2} \qquad (3)$$

$N_0$ is the equilibrium pore density for $\phi_m = 0$ and $q = (r_m/r^*)^2$, where $r_m = 0.8$ nm is the minimum energy radius at $\phi_m = 0$; and $r^* = 0.51$ nm is the minimum possible radius of a hydrophilic pore [16].

The pores are initially formed at $r_m$ and later expand in order to minimize the energy $W$ of the lipid bilayer (henceforth referred to as 'bilayer energy'). If at any point of time, there be $n$ number of pores with radii $r_j$, with $j=1,\ldots\ldots n$, then, the rate of change of radii is given by

$$\frac{dr_j}{dt} = -\frac{D_p}{k_B T}\frac{\partial W}{\partial r_j}, \quad j=1,2,\ldots\ldots,n \tag{4}$$

where $D_p = 5 \times 10^{-14}$ m$^2$ s$^{-1}$ is the diffusion coefficient for pore radius.
The bilayer energy depends on the repulsion of lipid heads, edge energy of pore perimeter, membrane tension and force due to the applied electric field.

The bilayer energy W is given by:

$$W = \sum_{j=1}^{n}\left\{\beta\left(\frac{r_0}{r_j}\right) + 2\pi r_j(\pi\kappa)\left(\frac{1}{h} - \psi c_0\right) - \pi\sigma_{eff}(A_p)r_j^2 + \int_0^{r_j} F(r_j, \phi_m)\,dr\right\} \tag{5}$$

In equation (5), the first term represents the static repulsion of the lipid heads, $\beta = 1.4 \times 10^{-19}$ J; the second represents edge energy of pore perimeter; the third term accounts for the effect of pores in the membrane tension, and the last term represents the contribution of the transmembrane potential.

The presence of cholesterol in the membrane has been incorporated in the second term in equation (5), which deals with the edge energy of the pore perimeter. This is a major difference between our model and those published earlier [16]. The presence of cholesterol is reflected in the change in line tension γ of the pore edge, when compared to the membrane without cholesterol. This is because of the negative spontaneous curvature ($c_0$ = -0.9 nm$^{-1}$) for cholesterol as found by Karatekin et al [17]. Thus, actual line tension of the membrane containing cholesterol is given by,

$$\gamma = \pi\kappa\left(\frac{1}{h} - \psi c_0\right) \tag{6}$$

where $\kappa = 2.7 \times 10^{-20}$ is the bending modulus of the membrane, $h$ = 5 nm (membrane thickness) and $\psi$ is the mole-fraction of cholesterol in the membrane in the rim around the pore-edge. In the present study, an average mammalian cell membrane containing a mole-fraction of cholesterol of ~20% (i.e, $\psi$ = 0.2) has been considered [18]. The variation in the cholesterol content is supposed to be a major factor contributing to differences in the membrane composition from one cell type to another. Hence, differences in the membrane properties may affect the electroporation behaviour of the cells. This may be correlated with the recent finding by Kanduser et al [19], who have demonstrated that different cell types having comparable cell geometries but different membrane properties (evaluated in terms of intrinsic membrane fluorescence) differed in electroporation behaviour when exposed to similar electric pulses.

The third term in equation (5) denotes the effective tension of the membrane, $\sigma_{eff}$ as the function of the combined area of pores

$$A_p = \sum_{j=1}^{n}\pi r_j^2$$

and is given by

$$\sigma_{eff}(A_p) = 2\sigma' - \frac{2\sigma' - \sigma_0}{(1 - A_p/A)^2} \tag{7}$$

where σ' is the energy per unit area of hydrocarbon–water interface and *A* is the total membrane surface area.

The fourth term in equation (5) accounts for contribution for the transmembrane potential, with the force acting on the inner surface of the torridal pores and where F is defined as:

$$F(r_p, \phi_m) = \frac{F_{max}}{1 + \frac{r_h}{r_p + r_t}} \phi_m^2 \tag{8}$$

where $F_{max} = 6.9 \times 10^{-10}$ N/V$^2$ is the maximum attainable force when $\phi_m = 1$V , $r_h = 0.97$nm, $r_t = 0.31$ nm [15]. As pores are formed they expand and this expansion affects the trans-membrane potential $\phi_m$.

For the surface distribution of pores, that is, to incorporate the effect of variation in pore population with *θ,* the cell has been assumed to be divided into segments or discs with *θ* varying as:
θ = 0 to π/8 (segment T1);  θ = π/8 to π/4 (segment T2); θ = π/4 to 3π/8 (segment T3); θ =  3π/8 to π/2 (segment T4) as shown in the figure 1(b). The remaining half of the cell has also been evaluated in the same fashion. For simplicity of numerical implementation, the mean value of *θ* for each segment has been considered as the numerical input (for *θ*) for that particular segment. This is also a major difference with our model and those published earlier [16].

To compute $\phi_m$ for each segment at each time step, it is necessary to choose a type of experimental preparation to model.  The transmembrane potential is given by

$$C\frac{d\phi_m}{dt} + \left(\frac{1}{R_s} + \frac{1}{R}\right)\phi_m + I_p = \frac{\phi_0}{R_s}, \tag{9}$$

$R_s$ is 100 ohm in series resistance of the experimental setup (Device for carrying out electroporation, buffer etc,) $R = R_m/A$ is the membrane resistance with $R_m = 0.523$ ohm-m$^2$ (assumed to be the same for all segments).
$I_p$ is the combined current through all pores at a particular segment and is given by

$$I_p = \sum_{j=1}^{n} i_p(r_j, \phi_m) = \sum_{j=1}^{n} \frac{\phi_m}{R_p + R_i} = \sum_{j=1}^{n} \frac{\phi_m}{\frac{h}{g\pi r_j^2} + \frac{1}{2gr_j}} \tag{10}$$

where h =  5 nm is the membrane thickness, g = 2 S/m is the conductivity of the solution. $R_p = h/(\pi g r_j^2)$ is the pore resistance and $R_i = 1/2 gr_j$ is the input resistance [16].  Total current through the cell has been calculated by adding up of the currents for the various segments.

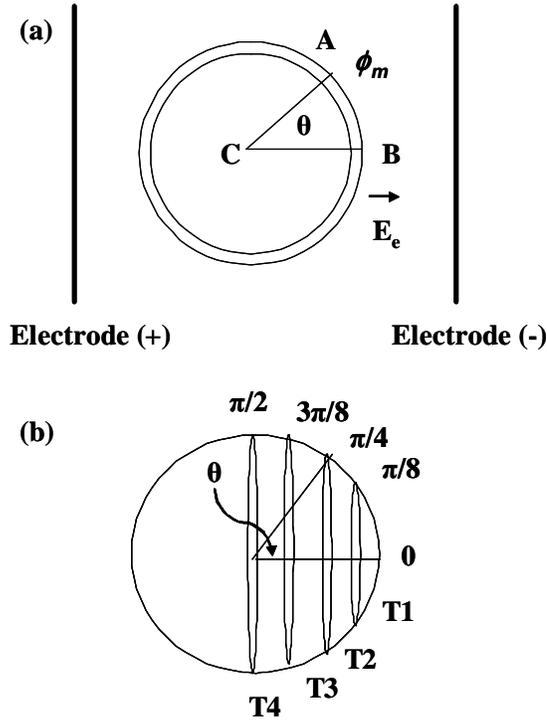

**Figure 1.** (a) Spherical cell in an applied external electric field $E_e$. (b) The cell is divided into segments as: $\theta = 0$ to $\pi/8$ (segment T1); $\theta = \pi/8$ to $\pi/4$ (segment T2); $\theta = \pi/4$ to $3\pi/8$ (segment T3); $\theta = 3\pi/8$ to $\pi/2$ (segment T4).

**3. Computational model**

A program has been developed in $C^{++}$ language for the dynamic simulation of cell electroporation. Simulations were carried out to obtain the numerical solutions for the equations.

For simulations, our model represents the pores in 2 populations as: i) small pores, whose radii congregate near the minimum energy radius ($r_m$), are accounted for by pore density $N(t)$, and ii) large pores with radii greater than $r_m$ are accounted by number of pores $n$ ($r_j$, $j = 1, 2, 3......, n$). It has been assumed that most of the hydrophilic pores are created at the minimum energy radius ($r_m = 0.8$ nm) [16].

Then, the transmembrane potential and all other electrical parameters like current are calculated separately for each of the four regions as shown by figure 1(b). Equation (9) has been used for calculating $\phi_m$, which is assumed to be constant throughout each region. The only difference is in the values of $\phi_0$ used for the different regions. $\phi_0$ is the maximum potential that can be developed across the membrane and is calculated using equation (1).

Thus, each half of the cell is treated as four independent regions as far as the electrical properties of the cell are concerned. But the mechanical properties like membrane surface tension, etc were considered the same for all the segments of the cell. Thus, the physical integrity of the cell is not lost.

The entire time duration of the simulation is divided into small time steps, each of duration $\Delta t = 0.2$ ns $= 2 \times 10^{-4}$ μs. Simulations were carried out for a total time duration of 40 us which is divided into i) time for application of the electric pulse $T_{ON} = 20$ μs and ii) post pulse duration, $T_{OFF}$ = 20 µs during which the membrane tries to rearrange. It has also been assumed that at time $t = 0$, assume $\phi_m = 0$, $N = N_0$ and $n = 0$. $N_0$ is the equilibrium pore density at $\phi_m = 0$ and depends on the rate of thermal fluctuations of lipid molecules in the membrane.

A single time step in the simulation for each region consists of the following calculations:
1. Spontaneous creation of pores according to equation (2) simultaneously in all the segments of the cell.
2. If the number of pores $N(t)$ x $A$ (membrane area for given region) >1, then $n$ is increased by an integer number. For example, if $N(t) \times A = 2.3$, then increase $n$ by 2 and decrease $N(t)$ by $2/A$.
3. Update radii ($r_j$) of $n$ pores according to equation (4). If any of these pores have new radii equal to minimum energy radius $r_m$ then value of $n$ is decreased by integer number of pores and make corresponding increase in the value of $N(t)$.
4. Membrane tension $\sigma_{eff}$ is calculated only once for each time step, according to the equation (6). This value is used while updating the pore radii according to equation (4).
5. The transmembrane potential, $\phi_m$ has to be calculated according to equation (9). Current through each individual pore is computed and pores from both populations are considered during this calculation using density of pores $N(t)$, and the radii of pores $r_j$, ($j = 1, 2, 3……, n$). Then total current $I_p$ is calculated using equation (10).
6. For each segment, the mean value of $\theta$ was considered as the numerical input for calculation of $\phi_m$. For example, for segment 1 ($\theta = 0$ to $\pi/8$), $\theta = \pi/16$ was considered at the input.
7. The numerical integration of the equations (2), (4) and (9) has been performed using the Runge-Kutta $4^{th}$ order method.
8. To incorporate the presence of cholesterol in the membrane, the numerical implementation of equation (4) has been performed using the value of bilayer energy $W$ as defined by the equation (5).
9. Values of $R_{avg}$ (arithmetic mean of all pore radii in a segment), $R_{max}$ (max pore radius in a segment), $\phi_m$ and pore-count are recorded in the data files.

Simulations were performed for two different values of applied electric field: i) Ee = 0.833 kV/cm (for comparing results with previous model proposed by[16]) and ii) $E_e$ = 1.0 kV/cm (for understanding the phenomenon of electroporation at higher applied electric field magnitudes). The program for simulations has been used on a Pentium 4 PC with Windows XP, 1.80 GHz., 256 MB RAM. The program generates data files ($\phi_m$ vs $t$, $R_{avg}$ vs $t$, $R_{max}$ vs $t$, Pore count vs $t$) for all segments.

**4. Results and discussion**
In this study, our model incorporates the effects of surface distribution of pores for a spherical cell of radius ($R_{cell}$= 10 µm) whose membrane consists of cholesterol (20% mole-fraction). Results for simulations performed with two different values of applied electric field, i) $E_e$ = 0.833 kV/cm and ii) $E_e$ = 1.0 kV/cm, are discussed below.

*4.1 Electroporation with $E_e$ = 0.833 kV/cm:*
The simulation of cell electroporation, in which electric pulses of 20 µs duration have been applied, is shown by Figure 2. The total time interval is divided into i) time for electric pulse application (duration 20 µs) and ii) a post-pulse duration of 20 µs. As shown by the figure 2(a), the transmembrane potential, $\phi_m$, for various segments of the cell increases to a maximum value and then falls to 0V at the end of the pulse. For segment T1, the value of $\phi_m$ attains a value of

~1.2 V and then stabilizes at 1.14 V. For the segment T2 and T3, $\phi_m$ attains constant values of 1.04 V and 0.69 V respectively through-out the pulse duration. The least value of $\phi_m$ is obtained for segment T4 (~0.24 V).

Figure 2(b) shows the variation of average pore radius $R_{avg}$ with time during the electroporation. For segment T1, $R_{avg}$ increases and stabilizes at 63.64 nm ≈ 64 nm. At the end of the pulse the value of $R_{avg}$ decreases and reaches ~15 nm (at t= 40 μs). For segment T2, the $R_{avg}$ increases to a maximum value of ~32 nm and falls sharply at the end of the pulse, indicating the resealing of pores. Fig 2(c) shows that for the segment T1, $R_{max}$ reaches a maximum value of ~82 nm and then gradually decreases to ~77 nm towards the end of the pulse duration. For segment T2, $R_{max}$ attained a value of ~57 nm initially and then, stabilizes at ~55 nm. The difference in values of $R_{max}$ and $R_{avg}$ is less, indicating that majority of the pores in each segment attain similar radii close to the $R_{avg}$ for that segment.

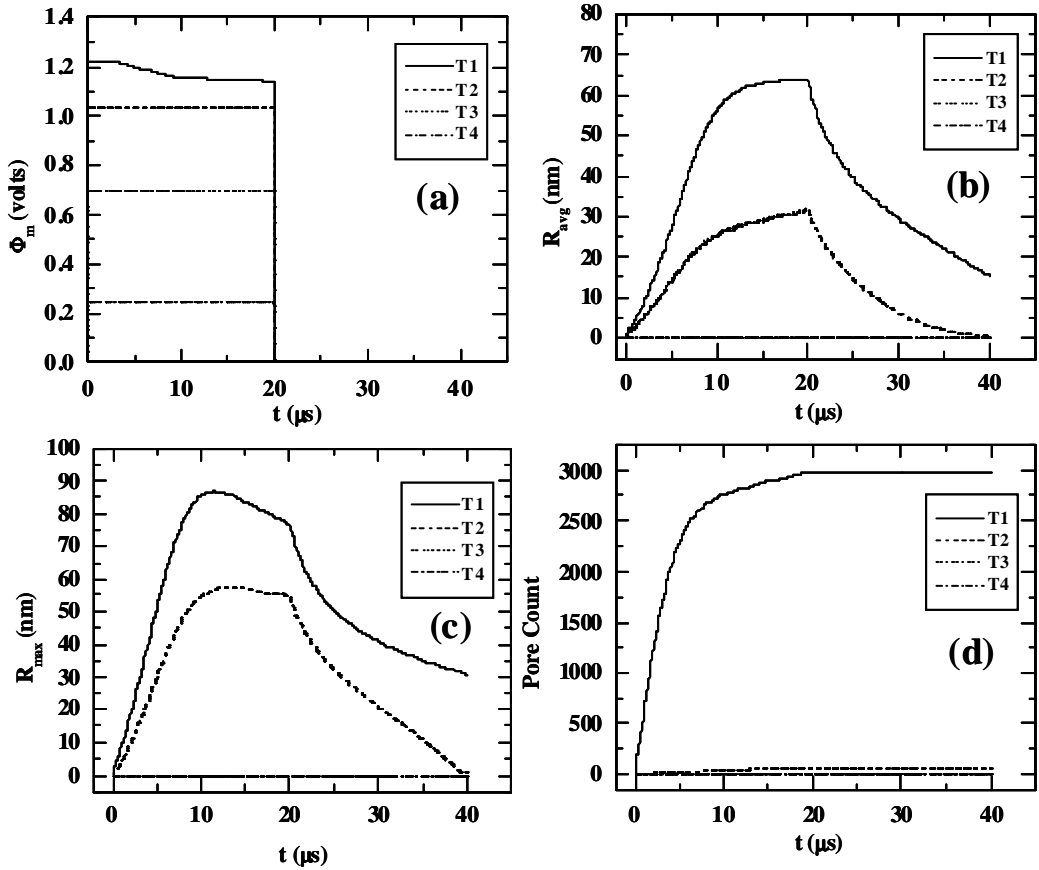

**Figure 2.** The figures (a), (b) and (c) represent the variation of transmembrane potential ($\phi_m$), average pore radius ($R_{avg}$) and maximum pore radius ($R_{max}$) with time respectively. Each figure shows 4 curves corresponding to 4 regions of cell placed in an electric field of 0.8333kV/cm: T1(θ = 0 to θ = π/8), T2(θ = π/8 to θ = π/4), T3(θ = π/4 to θ = 3π/8) and T4(θ = 3π/8 to θ = π/2). Figure (d) indicates the pore distribution for various segments.

The graph also shows that there is no significant pore formation in the segments T3 and T4 ($R_{avg}$ = 0 nm) (Fig 2(d)). The program also predicts that the number of pores formed in segments T1 and T2 are 2992 and 60 respectively, whereas there is no pore formation in the other segments.

Our model considers the effective electric fields in different segments of the cell surface. The total number of pores formed in a cell as per this model is far less as compared to that predicted by a model considering the cell as a uniform membrane patch (model of Smith et al). Our model predicts that most of the pores are formed in the segment T1, leading to a slight drop in the value of $\phi_m$ during the application of the pulse for that segment. Still a high value of $\phi_m$ is maintained during the entire pulse-duration leading to an increase in the pore size. Hence, the average radius for pores in this region reaches a maximum of ~64 nm, which is significantly greater (~1.9 times) than that predicted (~34 nm) by [16]. It is evident that the number of pores predicted by this model is lesser as compared to that predicted by [16].

*4.2 Electroporation with $E_e$ = 1.0 kV/cm:*
Simulations of cell electroporation have also been carried out for an applied field of 1 kV/cm and results are displayed by Figure 3. Figure 3(a) shows that for segment T1, $\phi_m$ increases abruptly to ~1.45 V with the application of the electric pulse and then stabilizes at ~0.84 V. On withdrawal of the pulse, $\phi_m$ decreases to 0 V. This initial spike is similar to that reported by Smith *et al* [16]. But for segment T2, $\phi_m$ increases to ~1.24 V and stabilizes at ~1.152 V till the end of the pulse and then falls to 0 V. For segments T3 and T4, $\phi_m$ attains constant values of 0.83 V and 0.29 V respectively.

Figure 3(b) shows the variation of $R_{avg}$ with time for the various segments of the cell. For segment T1, the radius of the pore increases abruptly and stabilizes at ~8 nm for the duration of the pulse. On withdrawal of the pulse, $R_{avg}$ falls to ~0.8 nm, indicating closing of the pores. For segment T2, the $R_{avg}$ is maximum (~ 23 nm). The value of $R_{avg}$ decreases on withdrawal of pulse. It is evident from the figure 3 that larger pores are generated in the segment T2 whereas smaller pores are formed in the segment T1. Figures 3 (b & c) show that for each segment values of $R_{max}$ and $R_{avg}$ are very close. This signifies that in each segment almost all pores attain the same radius, ~ $R_{avg}$ for that segment. Also, Figure 3(d) indicates that the pore population is more in segment T1 (349285) as compared to that in segment T2 (11,192). The graph also indicates that there is no significant pore formation in segments T3 and T4.

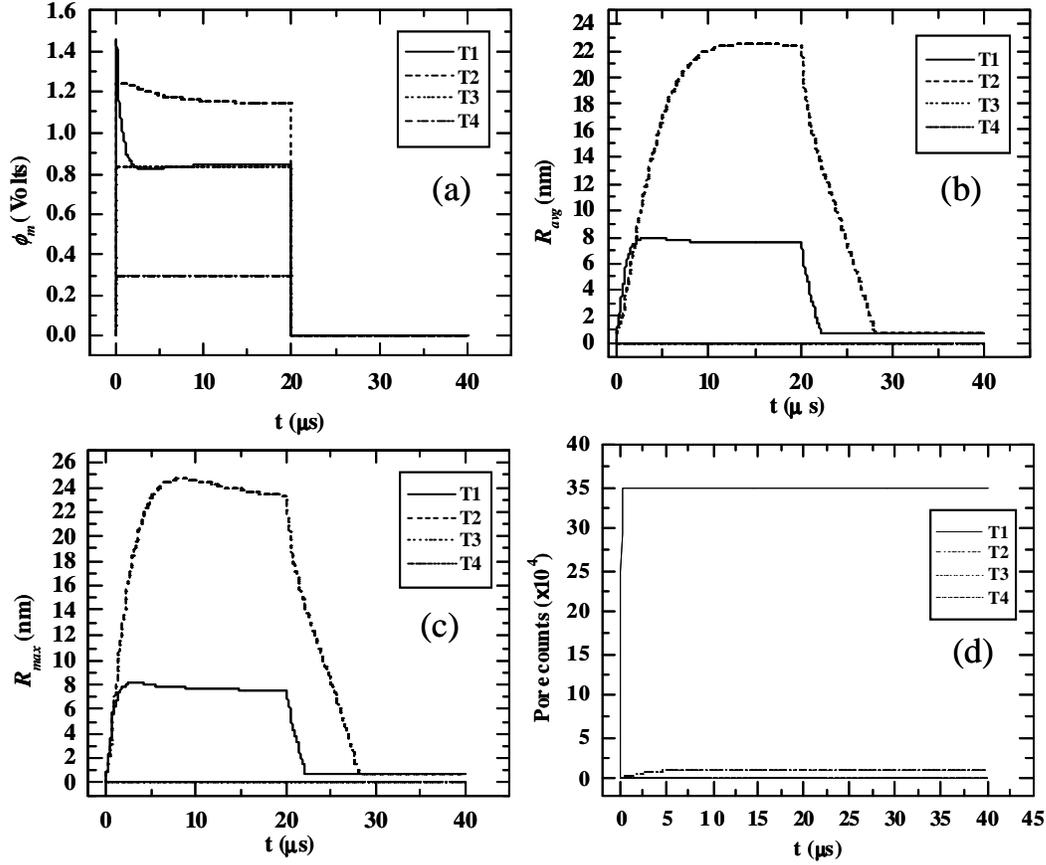

**Figure 3.** The figures (a), (b) and (c) represent the variation of transmembrane potential ($\phi_m$), average pore radius ($R_{avg}$) and maximum pore radius ($R_{max}$) with time, respectively. Each figure shows 4 curves corresponding to 4 regions of cell placed in an electric field of 1 kV/cm: T1($\theta = 0$ to $\theta = \pi/8$), T2($\theta = \pi/8$ to $\theta = \pi/4$), T3($\theta = \pi/4$ to $\theta = 3\pi/8$) and T4($\theta = 3\pi/8$ to $\theta = \pi/2$). Figure (d) indicates the segment-wise distribution of pores over the cell surface.

This model predicts that for segment T1, the transmembrane potential $\phi_m$ abruptly increases to ~1.45 V at the onset of the pulse. This leads to the creation of large number of pores within a very short time (~2 μs). This abrupt increase in pore population increases the permeability (and hence the conductivity) of the membrane thereby reducing $\phi_m$ to ~0.8 V for the rest of the pulse duration. Such low value of $\phi_m$ does not favour further expansion of pores. This is evident from a constant value of $R_{avg}$ maintained during the rest of the pulse.

However, for segment T2, $\phi_m$ attains an initial value of only ~1.24 V leading to the formation of 11,192 pores, which is much less as compared to that in segment T1. This causes a higher transmembrane potential ($\phi_m$ ~1.15 V) to be maintained for the entire pulse duration. Consequently, the pores expand to larger radii, with the $R_{avg}$ being ~23 nm. Such large pores cannot be predicted by models that consider uniform membrane poration [16] for higher values of electric fields viz $E_e$~1 kV/cm or more. The simultaneous existence of large pores (in segment T2) and a large number of small pores (in segment T1) indicate the possibility of effective electroporation with the chosen electroporation parameters.

*4.3 Future prospect*

The present model describes the electroporation of a cell membrane containing cholesterol (~20%). The model predicts the occurrence of significant poration in two segments T1 and T2 of the cell with lower values of $\theta$ (i.e, regions with $\theta = 0$ to $\theta = \pi/4$) and no poration in the other segments. This is in agreement with the experimental results published earlier [20]. However, the model has not been extended to predict the uptake of DNA. This may be taken up as a future project.

Also, the cholesterol content of mammalian cells depends on the type of the biological cell. General case with (20% mole fraction) cholesterol content has been considered for simulation in this model. The dependence of nature of pore formation (poration) on cholesterol content of mammalian cells (all possible values of $\psi$) also can be taken up as a future course of study.

**5. Conclusion**

In this article a method for numerical implementation (simulations) of electroporation of a mammalian cell (with cholesterol containing membrane) has been developed. Based on the model we have analyzed the variation of transmembrane potential $\phi_m$, size of pores $R_{avg}$, $R_{max}$ and pore-count with time for different segments of a biological cell during electroporation. For simulation of electroporation phenomenon with a more realistic approach, two modifications were made to the existing model [16]: a) the incorporation of presence of cholesterol, and b) surface distribution of pores (i.e the variation of number of pores with $\theta$).

The results are summarized below:
   i)   The consideration of cholesterol in membrane has effectively increased the average pore radius as compared to that predicted for a cholesterol-free membrane [16] for the same magnitude of applied electric field.
   ii)  This model quantitatively predicts the surface distribution of pores i.e., variation in number of pores with $\theta$. The results clearly show greater poration in polar regions (regions with $\theta = 0$ to $\theta = \pi/4$) as compared to the remaining parts of the cell. Hence, the total number of pores predicted for the whole cell is different as compared to that predicted by other models [16].
   iii) The model can predict enhanced poration and effective electroporation for higher magnitudes of applied electric fields ($E_e$ ~1 kV/cm onwards).

This model can be a useful tool for predicting a more realistic nature of poration in electroporation of mammalian cells at moderate and higher magnitudes of applied electric fields. It can also help researchers in understanding the dependence of electroporation on cholesterol content of cell membranes.

**Acknowledgements**
The authors would like to thank Dr. S. H. Sanghvi and Dr. K.P. Mishra, Ex-Scientists, Radiation Biology and Health Sciences Division, BARC, Mumbai, India for meaningful discussions. This work has been supported by the School of Basic Medical Sciences at the Department of Physics, University of Pune, Pune, India.